© Indian Academy of Sciences

# FRW model with two-fluid source in fractal cosmology

D D PAWAR[1] ,*, D K RAUT[2] and W D PATIL[3]

[1]School of Mathematical Sciences, Swami Ramanand Teerth Marathwada University, Nanded 431 606, India
[2]Department of Mathematics, Shivaji Mahavidyalaya, Renapur, Latur 413 527, India
[3]Department of Applied Mathematics, A.C. Patil College of Engineering, Navi Mumbai 410 210, India
*Corresponding author. E-mail: dypawar@yahoo.com

**Abstract.** The present paper deals with flat Friedmann–Robertson–Walker (FRW) model with two-fluid sources in fractal cosmology. One fluid in this model represents the Universe's matter content, while the other is a radiation field that models the cosmic microwave background. To obtain the deterministic model, we used the pressure–density relationship for matter via the gamma law equation of state $p_m = (\gamma - 1)\rho_m$, $1 \leq \gamma \leq 2$. The solutions of fractal field equations are obtained in terms of Kummer's confluent hypergeometric function of the first kind. Some physical parameters of the models are obtained and their behaviour is discussed in detail using graphs.

## 1. Introduction

The Friedmann–Robertson–Walker (FRW) model, which is homogeneous and isotropic, is thought to be a good approximation of the early and current stages of the Universe. The 2.73K CMBR has attracted many researchers to study the FRW model with two fluids [1–3]. In literature it is found that rather than using two sequences of a single fluid model, a cosmic evolution based on two-fluid models is preferable. In two-fluid models, one fluid dominates the other, and both fluids are assumed to be present throughout the cosmic evolution. Such models are used to describe the evolution of the Universe from a radiation-dominated phase to a matter-dominated phase. The radiation and matter fluid models are cosmologically important. Their applications to cosmology are inspired by observations [4] that recommend that the radiation frame and therefore the matter frame of the Universe might not coincide.

In this work, we consider a homogeneous and isotropic FRW Universe with two fluid sources within the framework of fractal cosmology. Many researchers are interested in studying the two-fluid cosmological models in general relativity as well as in alternative theories of gravity. Goswami and Pradhan [5] have recently studied a two-fluid cosmological model in a Bianchi type-I Universe. Rao *et al* [6] investigated homogeneous Kantowski–Sachs two-fluid rotating cosmological model in Brans–Dicke theory of gravitation. Chimento [7] investigated interacting fluids that produce identical, dual and phantom cosmologies. In general relativity, Samanta [8] examined two-fluid anisotropic Bianchi type-III cosmological model with variable $G$ and $\Lambda$. Pawar and Solanke [9] studied the two-fluid dark energy cosmological model in the Saez Ballester theory of gravity. Adhav *et al* [10] studied the two-fluid cosmological models in Bianchi type-III space–time. In the theory of general relativity and the Lyra manifold, a bulk viscous fluid with magnetised string and plane symmetric string cosmological models has been studied by Pawar *et al* [11–13]. Mishra *et al* [14] studied the cosmological model of an accelerating dark energy in two fluids, with a mixed scale factor. Katore and Hatkar [15] have investigated FRW Universe in the framework of $f(R)$ gravity with two co-moving perfect fluids as the source of gravitational fluid. Singh *et al* [16] have explored rotating anisotropic two-fluid Universe coupled with radiation and a scalar field. Tiwari *et al* [17,18] have analysed EoS parameter for dark energy in FRW and Bianchi type-III space–time filled with dark energy and barotropic fluid. Pawar and Dagwal [19] studied two fluid sources in $F(T)$ theory of gravitation. Mishra *et al* [20] studied stability of the two fluid cosmological models. Hasmani and Pandya [21] investigated cosmological models for the static spherically symmetric space–time with charged anisotropic fluid distribution in the context of Rosen's bimetric general relativity (BGR).

Recently, fractals have been used to study the Universe. Mandelbrot [22] developed the concept of fractals and proposed that the standard cosmological principle be replaced by a conditional cosmological principle. Pietronero [23] explained the fractal distribution of galaxies to precisely fit astronomical data. On the scale of relativity, Laurent [24] proposed the fractal nature of space–time. Eternally existing chaotic inflationary model of the Universe was investigated by Linde [25]. Calcagni [26] investigated the flat FRW model within a fractal framework. In fractal theory of gravitation, Mustafa Salti *et al* [27] investigated the extended versions of holographic and Ricci dark energy models. Karami *et al* [28] studied the flat fractal FRW Universe in dark energy and dark matter. In fractal cosmology, Hosseienkhani *et al* [29] investigated the anisotropic behaviour of dark energy. Aly and Selim [30] studied $f(T)$-modified gravity model for the dark energy in fractal Universe. Recently, Pawar *et al* [31] studied FRW space–time within the domain wall in the framework of fractal cosmology.

Motivated by the aforementioned studies, the current paper attempts to investigate the flat FRW model with two fluids within the framework of fractal cosmology. We assumed the fractal parameter $\omega \neq 0$ and the power law form of the fractal function $v$ to obtain the solutions of fractal field equations. The analysis was carried out in the UV regime with no cosmological constant. In §2, we have presented the formalism of fractal gravity. In §3, we obtained the fractal field equations. Solutions of field equations under the influence of fractal cosmology are derived in §4. Section 5 deals with a brief discussion of the results. Finally, §6 highlights our conclusions.

## 2. Formalism of fractal gravity

Calcagni [26] proposed a fractal cosmological model for power counting renormalisable field theory living in fractal space–time. The action in this model is Lorentz invariant and equipped with Lebesue–Stieltjes measure. The total action of Einstein's gravity in fractal space–time, assuming matter is minimally coupled with gravity, can be written as [26]

$$S = S_g + S_m, \tag{1}$$

where $S_g$ is the gravitational part of the action

$$S_g = \frac{M^2}{2} \int \sqrt{-g}(R - 2\Lambda - \omega \partial_a v \, \partial^a v) \, d\xi(x) \tag{2}$$

and $S_m$ is the matter part of the action

$$S_m = \int \sqrt{-g} \, L_m \, d\xi(x). \tag{3}$$

Here $M^2 = 1/8\pi G$, $\Lambda$, $R$, $L_m$ and $g$ describe the reduced Planck mass, cosmological constant, Ricci curvature scalar, Lagrangian density of the matter part of action and determinant of dimensionless metric $g_{ab}$ respectively. The quantities $\omega$ and $v$ are fractal parameter and fractal function respectively. The standard measure $d^4x$ in the action is replaced by Lebesgue–Stieltjes measure $d\xi(x)$ and $d\xi(x) = (d^4x)v(x)$.

The Einstein's equations in fractal Universe are derived by varying total action (1) with respect to the metric tensor $g_{ab}$

$$R_{ab} - \frac{1}{2}g_{ab}(R - 2\Lambda) + g_{ab}\frac{\Box v}{v} - \frac{\nabla_a \nabla_b v}{v}$$
$$+ \left(\frac{1}{2}g_{ab}\partial_\sigma v \, \partial^\sigma v - \partial_a v \, \partial_b v\right) = 8\pi G \, T_{ab}, \tag{4}$$

where $T_{ab}$ is the energy–momentum tensor given by

$$T_{ab} = \frac{-2}{\sqrt{-g}}\frac{\delta s_m}{\delta g^{ab}} \tag{5}$$

and $\Box = \nabla^\mu \nabla_\mu$ is the covariant derivative.

## 3. Metric and the field equations in fractal cosmology

Consider the FRW space–time in the form

$$ds^2 = -dt^2 + S^2(t)$$
$$\times \left(\frac{dr^2}{1 - kr^2} + r^2(d\theta^2 + \sin^2\theta \, d\phi^2)\right), \tag{6}$$

where the scale factor $S(t)$ is a function of cosmic time $t$, $k$ is a constant which denotes the spatial curvature and $k = 1, 0, -1$ represents closed, flat and open FRW Universe, respectively.

The spatial volume ($V$), Hubble parameter ($H$) and expansion scalar ($\theta$) for metric (6) are respectively given by

$$V = S^3, \quad H = \frac{\dot{S}}{S}, \quad \theta = 3\frac{\dot{S}}{S}, \tag{7}$$

where the overhead dot denotes the derivative with respect to cosmic time ($t$).

The deceleration parameter is defined as

$$q = \frac{-S\ddot{S}}{\dot{S}^2}. \tag{8}$$

The energy–momentum tensor for the two-fluid source has the form

$$T_{ij} = \overset{(m)}{T}_{ij} + \overset{(r)}{T}_{ij}, \tag{9}$$

where $\overset{(m)}{T}_{ij}$ is the energy–momentum tensor for the matter field with energy density $\rho_m$ and pressure $p_m$ and $\overset{(r)}{T}_{ij}$ is the energy–momentum tensor for the radiation field having energy density $\rho_r$ and pressure $p_r = \rho_r/3$, which are respectively given by

$$\overset{(m)}{T}_{ij} = (p_m + \rho_m)\overset{(m)}{u_i}\overset{(m)}{u_j} + p_m g_{ij} \tag{10}$$

$$\overset{(r)}{T}_{ij} = \frac{4}{3}\,\rho_r\overset{(m)}{u_i}\overset{(m)}{u_j} + \frac{1}{3}\rho_r g_{ij}. \tag{11}$$

The four-velocity vectors are $\overset{(m)}{u_i} = (1, 0, 0, 0)$ and $\overset{(r)}{u_i} = (1, 0, 0, 0)$ with $g^{ij}\overset{(m)}{u_i}\overset{(m)}{u_j} = -1$ and $g^{ij}\overset{(r)}{u_i}\overset{(r)}{u_j} = -1$.

Using eqs (10) and (11) in eq. (4) one can obtain Friedmann equations in the fractal Universe as

$$H^2 + H\frac{\dot{v}}{v} - \frac{1}{6}\omega\,\dot{v}^2 + \frac{k}{S^2} = \frac{8\pi G}{3}(\rho_m + \rho_r) + \frac{\Lambda}{3} \tag{12}$$

$$H^2 + \dot{H} - H\frac{\dot{v}}{v} - \frac{1}{2}\frac{\Box v}{v} + \frac{1}{3}\omega\,\dot{v}^2$$
$$= \frac{-8\pi G}{6}(3p_m + \rho_m + 2\rho_r) + \frac{\Lambda}{3}. \tag{13}$$

It is worth noting that putting $v = $ constant and $\omega = 0$ in eqs (12) and (13) yields standard Friedmann equations in Einstein's general theory of relativity. The purely gravitational constraint in fractal Universe is given by

$$3H^2 + \dot{H} + H\frac{\dot{v}}{v} + \frac{\Box v}{v} + \frac{2k}{S^2} - \omega(v\Box v - \dot{v}^2) = 0. \tag{14}$$

Taking time-like fractal profile of $v$ (power-law form) is $v = t^{-\beta}$, where $\beta$ being a positive constant, one can obtain

$$\frac{\Box v}{v} = \frac{\beta}{t}\left(3H - \frac{1+\beta}{t}\right). \tag{15}$$

The power-law form of the fractal function ($v$) is consistent with recent observations, particularly for describing the early era of the evolution of the Universe. Karami *et al* [28], Aly and Selim [30] and Ghaffari *et al* [32] are some of the researchers who have studied cosmological models in the framework of fractal cosmology using the power-law form of the fractal function and they found that it is consistent with the recent observations of accelerated expansion of the Universe. If $v = $ constant, the theory's predictions approach those of Einstein's standard theory of relativity. Without cosmological constant ($\Lambda = 0$), in UV regime ($\beta = 2$), units are to be chosen such that $8\pi G = 1$. For the flat Universe ($k = 0$), eqs (12)–(14) take the following form:

$$H^2 - \frac{2}{t}H - \frac{2}{3}\frac{\omega}{t^6} = \frac{1}{3}(\rho_m + \rho_r) \tag{16}$$

$$H^2 + \dot{H} - \frac{1}{t}H + \frac{3}{t^2} + \frac{4}{3}\frac{\omega}{t^6}$$
$$= \frac{-1}{6}(3p_m + \rho_m + 2\rho_r) \tag{17}$$

$$3H^2 + \dot{H} + \left(2 + \frac{3\omega}{t^4}\right)\frac{2}{t}H = \frac{6}{t^2} + \frac{10\omega}{t^6}. \tag{18}$$

## 4. Solutions of field equations and some physical parameters of the model

To solve the field equations, we assume that the fractal parameter $\omega \neq 0$ in eq. (18) and on solving it, we get the scale factor ($S$) [26] as

$$S = \frac{1}{t^2}\,\phi\left(\frac{11}{4}, \frac{13}{4}, \frac{3\omega}{2t^4}\right)^{1/3}, \tag{19}$$

where $\phi$ is Kummer's confluent hypergeometric function of the first kind and it is given as

$$\phi(a, b, x) = \frac{\Gamma b}{\Gamma a}\sum_{n=0}^{+\infty}\frac{\Gamma(a+n)}{\Gamma(b+n)}\frac{x^n}{n!}. \tag{20}$$

Applying eq. (19) in eq. (7), the spatial volume ($V$), Hubble parameter ($H$) and expansion scalar ($\theta$) of the model are given by.

$$V = \frac{1}{t^6}\,\phi\left(\frac{11}{4}, \frac{13}{4}, \frac{3\omega}{2t^4}\right) \tag{21}$$

$$H = \frac{-2}{t} - \frac{2\omega}{t^6} + \frac{4\omega}{13t^5}\frac{\phi\left(\frac{11}{4}, \frac{17}{4}, \frac{3\omega}{2t^4}\right)}{\phi\left(\frac{11}{4}, \frac{13}{4}, \frac{3\omega}{2t^4}\right)} \tag{22}$$

$$\theta = \frac{-6}{t} - \frac{6\omega}{t^6} + \frac{12\omega}{13t^5}\frac{\phi\left(\frac{11}{4}, \frac{17}{4}, \frac{3\omega}{2t^4}\right)}{\phi\left(\frac{11}{4}, \frac{13}{4}, \frac{3\omega}{2t^4}\right)}. \tag{23}$$

From eqs (19) and (8), the deceleration parameter of the model is obtained as

$$q = \frac{-N}{D}, \tag{24}$$

where

$$N = 169(3t^4 + 4\omega)(t^4 + \omega)\phi\left(\frac{11}{4}, \frac{13}{4}, \frac{3\omega}{2t^4}\right)^2$$
$$+ 52\omega(2t^4 + \omega)\phi\left(\frac{11}{4}, \frac{17}{4}, \frac{3\omega}{2t^4}\right)$$
$$\times \phi\left(\frac{11}{4}, \frac{13}{4}, \frac{3\omega}{2t^4}\right)$$
$$- 16\omega^2\phi\left(\frac{11}{4}, \frac{17}{4}, \frac{3\omega}{2t^4}\right)^2 \tag{25}$$

$$D = 338(t^4 + \omega)^2 \phi\left(\frac{11}{4}, \frac{13}{4}, \frac{3\omega}{2t^4}\right)^2$$
$$- 104\omega(t^4 + \omega)\phi\left(\frac{11}{4}, \frac{17}{4}, \frac{3\omega}{2t^4}\right)$$
$$\times F\left(\frac{11}{4}, \frac{13}{4}, \frac{3\omega}{2t^4}\right)$$
$$- 8\omega^2 \phi\left(\frac{11}{4}, \frac{17}{4}, \frac{3\omega}{2t^4}\right)^2. \tag{26}$$

Equations (16) and (17) are two equations in four unknowns. Hence, to solve it, we consider the relation between pressure and energy density of the matter field through gamma law equation of state which is given by

$$p_m = (\gamma - 1)\rho_m, \quad 1 \le \gamma \le 2. \tag{27}$$

Solving eqs (16) and (17) with the help of eq. (27), we get

$$p_m = \frac{-2(\gamma - 1)}{3\gamma - 4}\left[\left(\frac{57t^8 + 86\omega t^4 + 24\omega^2}{t^{10}}\right)\right.$$
$$- \frac{12\omega}{13t^{10}}(3t^4 + 2\omega)\frac{\phi\left(\frac{11}{4}, \frac{17}{4}, \frac{3\omega}{2t^4}\right)}{\phi\left(\frac{11}{4}, \frac{13}{4}, \frac{3\omega}{2t^4}\right)}$$
$$\left. - \frac{48\omega^2}{13^2 t^{10}}\frac{\phi\left(\frac{11}{4}, \frac{17}{4}, \frac{3\omega}{2t^4}\right)^2}{\phi\left(\frac{11}{4}, \frac{13}{4}, \frac{3\omega}{2t^4}\right)^2}\right] \tag{28}$$

$$\rho_m = \frac{-2}{3\gamma - 4}\left[\left(\frac{57t^8 + 86\omega t^4 + 24\omega^2}{t^{10}}\right)\right.$$
$$- \frac{12\omega}{13t^{10}}(3t^4 + 2\omega)\frac{\phi\left(\frac{11}{4}, \frac{17}{4}, \frac{3\omega}{2t^4}\right)}{\phi\left(\frac{11}{4}, \frac{13}{4}, \frac{3\omega}{2t^4}\right)}$$
$$\left. - \frac{48\omega^2}{13^2 t^{10}}\frac{\phi\left(\frac{11}{4}, \frac{17}{4}, \frac{3\omega}{2t^4}\right)^2}{\phi\left(\frac{11}{4}, \frac{13}{4}, \frac{3\omega}{2t^4}\right)^2}\right] \tag{29}$$

$$\rho_r = \frac{6}{3\gamma - 4}\left(\frac{3(1 + 4\gamma)}{t^2} + \frac{\omega(6 + 17\gamma)}{t^6} + \frac{6\omega^2\gamma}{t^{10}}\right)$$
$$+ \frac{6}{3\gamma - 4}\left(\frac{12\omega(\gamma - 1)}{13t^{10}}(3t^4 + 2\omega)\right.$$
$$\times \frac{\phi\left(\frac{11}{4}, \frac{17}{4}, \frac{3\omega}{2t^4}\right)}{\phi\left(\frac{11}{4}, \frac{13}{4}, \frac{3\omega}{2t^4}\right)}$$

$$\left. - \frac{24\omega^2\gamma}{13^2 t^{10}}(\gamma - 2)\frac{\phi\left(\frac{11}{4}, \frac{17}{4}, \frac{3\omega}{2t^4}\right)^2}{\phi\left(\frac{11}{4}, \frac{13}{4}, \frac{3\omega}{2t^4}\right)^2}\right). \tag{30}$$

The total energy density is given by

$$\rho = \frac{t^8 + 12\omega^2 + 34\omega t^4}{t^{10}}$$
$$- \frac{24\omega^2}{13t^{10}}(3t^4 + 2\omega)\frac{\phi\left(\frac{11}{4}, \frac{17}{4}, \frac{3\omega}{2t^4}\right)}{\phi\left(\frac{11}{4}, \frac{13}{4}, \frac{3\omega}{2t^4}\right)}$$
$$+ \frac{48\omega^2}{13^2 t^{10}}\frac{\phi\left(\frac{11}{4}, \frac{17}{4}, \frac{3\omega}{2t^4}\right)^2}{\phi\left(\frac{11}{4}, \frac{13}{4}, \frac{3\omega}{2t^4}\right)^2}. \tag{31}$$

The required density parameters are

$$\Omega_m = \frac{-1}{3\gamma - 4}\left(\frac{I}{II}\right), \tag{32}$$

where

$$I = 169(72t^8 + 86\omega t^4 + 24\omega^2)\phi\left(\frac{11}{4}, \frac{13}{4}, \frac{3\omega}{2t^4}\right)^2$$
$$- 48\omega^2 \phi\left(\frac{11}{4}, \frac{17}{4}, \frac{3\omega}{2t^4}\right)^2$$
$$- 156(3t^4 + 2\omega)\phi\left(\frac{11}{4}, \frac{17}{4}, \frac{3\omega}{2t^4}\right)\phi$$
$$\times \left(\frac{11}{4}, \frac{13}{4}, \frac{3\omega}{2t^4}\right) \tag{33}$$

$$II = 169\left(6t^8 + 12\omega t^4 + 6\omega^2\right)\phi\left(\frac{11}{4}, \frac{13}{4}, \frac{3\omega}{2t^4}\right)^2$$
$$+ 24\omega^2 \phi\left(\frac{11}{4}, \frac{17}{4}, \frac{3\omega}{2t^4}\right)^2$$
$$- 312(t^4 + \omega)\phi\left(\frac{11}{4}, \frac{17}{4}, \frac{3\omega}{2t^4}\right)\phi$$
$$\times \left(\frac{11}{4}, \frac{13}{4}, \frac{3\omega}{2t^4}\right) \tag{34}$$

$$\Omega_r = \frac{-1}{3\gamma - 4}\left(\frac{A}{B}\right), \tag{35}$$

where

$$A = 169\left(-6t^8(1 + 4\gamma)\right.$$
$$- 3\omega t^4(6 + 17\gamma) - 18\omega^2\gamma\right)\phi\left(\frac{11}{4}, \frac{13}{4}, \frac{3\omega}{2t^4}\right)^2$$
$$+ 72\omega^2(2 - \gamma)\phi\left(\frac{11}{4}, \frac{17}{4}, \frac{3\omega}{2t^4}\right)^2$$
$$+ 156(3t^4 + 2\omega)(\gamma - 1)\phi\left(\frac{11}{4}, \frac{17}{4}, \frac{3\omega}{2t^4}\right)\phi$$

$$\times \left( \frac{11}{4}, \frac{13}{4}, \frac{3\omega}{2t^4} \right) \qquad (36)$$

$$B = 169 \left( 6t^8 + 12\omega t^4 + 6\omega^2 \right) \phi \left( \frac{11}{4}, \frac{13}{4}, \frac{3\omega}{2t^4} \right)^2$$
$$+ 24\omega^2 \phi \left( \frac{11}{4}, \frac{17}{4}, \frac{3\omega}{2t^4} \right)^2$$
$$- 312(t^4 + \omega)\phi \left( \frac{11}{4}, \frac{17}{4}, \frac{3\omega}{2t^4} \right) \phi$$
$$\times \left( \frac{11}{4}, \frac{13}{4}, \frac{3\omega}{2t^4} \right). \qquad (37)$$

At early times we distinguish the result for positive and negative values of fractal parameter $\omega$ and using the asymptotic form of Kummer's confluent hypergeometric function of the first kind

when $x \to \infty$

$$\phi(a, b, x) \sim \frac{\Gamma b}{\Gamma a} e^x x^{a-b}$$
$$+ e^x x^{a-b} \sum_{r=1}^{r=N} \frac{\Gamma(b-a+r)}{\Gamma(b-a)} \frac{\Gamma(b-a)}{\Gamma r} \frac{(-x)^{-r}}{r!}, \quad (38)$$

when $x \to -\infty$

$$\phi(a, b, x) \sim \frac{\Gamma b}{\Gamma(b-a)} (-x)^{-a}$$
$$+ (-x)^{-a} \sum_{r=1}^{r=N} \frac{\Gamma(a+r)}{\Gamma a} \frac{\Gamma(b)}{\Gamma(b-a-r)} \frac{(x)^{-r}}{r!} \quad (39)$$

and when $x \to 0$

$$\phi(a, b, x) \sim 1 \qquad (40)$$

we have the following different models.

### 4.1 *Model*-I: *When $t \to 0$ and $\omega > 0$*

Using eq. (38) the spatial volume, deceleration parameter, pressure and density for matter, density for radiation, total density and density parameters for matter and radiation of the model reduce to

$$V \sim \frac{e^{\frac{3\omega}{2t^4}}}{t^4} \qquad (41)$$

$$q \sim -1 - \frac{5t^4}{2\omega} \qquad (42)$$

$$p_m \sim \frac{-48\omega^2}{t^{10}} \left( \frac{\gamma - 1}{3\gamma - 4} \right) \qquad (43)$$

$$\rho_m \sim \frac{-48\omega^2}{t^{10}} \left( \frac{1}{3\gamma - 4} \right) \qquad (44)$$

$$\rho_r \sim \frac{36\omega^2}{t^{10}} \left( \frac{\gamma}{3\gamma - 4} \right) \qquad (45)$$

$$\rho \sim \frac{12\omega^2}{t^{10}} \qquad (46)$$

$$\Omega_m \sim \frac{-4}{3\gamma - 4} \qquad (47)$$

$$\Omega_r \sim \frac{3\gamma}{3\gamma - 4}. \qquad (48)$$

### 4.2 *Model*-II: *When $t \to 0$ and $\omega < 0$*

Using eq. (39) the spatial volume, deceleration parameter, pressure and density for matter, density for radiation, total density and density parameters for matter and radiation of the model reduce to

$$V \sim t^5 \qquad (49)$$

$$q \sim \frac{-2}{5} \qquad (50)$$

$$p_m \sim \frac{-136\omega}{t^6} \left( \frac{\gamma - 1}{3\gamma - 4} \right) \qquad (51)$$

$$\rho_m \sim \frac{-136\omega}{t^6} \left( \frac{1}{3\gamma - 4} \right) \qquad (52)$$

$$\rho_r \sim \frac{-6\omega}{t^6} \left( \frac{\gamma - 24}{3\gamma - 4} \right) \qquad (53)$$

$$\rho \sim \frac{-2\omega}{t^6} \qquad (54)$$

$$\Omega_m \sim \frac{-34\omega}{3t^4} \left( \frac{1}{3\gamma - 4} \right) \qquad (55)$$

$$\Omega_r \sim \frac{\omega}{2t^4} \left( \frac{24 - \gamma}{3\gamma - 4} \right). \qquad (56)$$

### 4.3 *Model*-III: *When $t \to \infty$ and $\omega > 0$*

Using eq. (40) the spatial volume, deceleration parameter, pressure and density for matter, density for radiation, total density and density parameters for matter and radiation of the model reduce to

$$V \sim \frac{1}{t^6} \qquad (57)$$

$$q \sim \frac{-3}{2} \qquad (58)$$

$$p_m \sim \frac{-114}{t^2} \left( \frac{\gamma - 1}{3\gamma - 4} \right) \qquad (59)$$

$$\rho_m \sim \frac{-114}{t^2} \left( \frac{1}{3\gamma - 4} \right) \qquad (60)$$

$$\rho_r \sim \frac{18}{t^2} \left( \frac{(1 + 4\gamma)}{3\gamma - 4} \right) \qquad (61)$$

$$\rho \sim \frac{24}{t^2} \qquad (62)$$

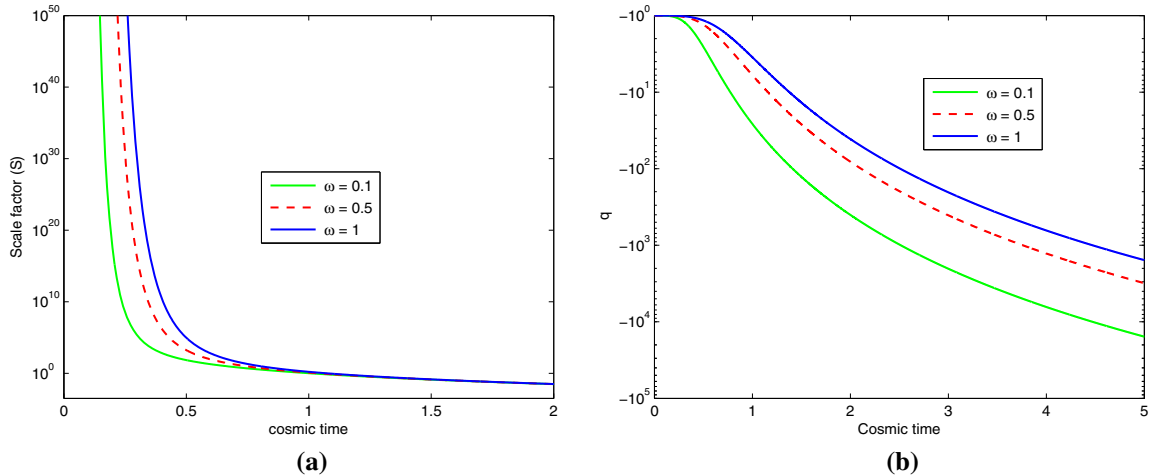

**Figure 1.** Variation of (**a**) scale factor ($S$) and (**b**) deceleration parameter ($q$) with cosmic time.

$$\Omega_m \sim \frac{-19}{2(3\gamma - 4)} \tag{63}$$

$$\Omega_r \sim \frac{18(1 + 4\gamma)}{3\gamma - 4}. \tag{64}$$

## 5. Discussion

In this section, we shall discuss the behaviour of physical parameters of the obtained models in detail. The spatial volume ($V$), Hubble parameter ($H$), expansion scalar ($\theta$), deceleration parameter ($q$), pressure for matter ($p_m$), energy density for matter ($\rho_m$), energy density for radiation ($\rho_r$) and total energy density ($\rho$) are obtained in the framework of fractal cosmology. It is found that all these parameters are functions of cosmic time $t$. Also, we discuss the behaviour of physical parameters using graphs. Graphs are drawn for a particular value of the fractal parameter $\omega$.

### 5.1 *The physical behaviour at early time ($t \to 0$) and $\omega > 0$*

The scale factor ($S$) of the current model is a decreasing positive function of time as shown in figure 1a. It has a very large initial value that decreases with time and approaches zero at a later time. It can be seen from eq. (41) that the volume of the model is infinite at $t = 0$, decreases with the passage of time and tends to zero as $t \to \infty$. The deceleration parameter $q$ plays an important role in the description of the late-time cosmic dynamics. A negative value of $q$ implies an accelerating Universe and a positive value of $q$ implies a decelerating Universe. In figure 1b, the variation of $q$ against cosmic time is shown for $\omega = 0.1, 0.5$ and $1$. It is observed that $q$ remains negative during the evolution of the Universe and at the initial moment $q \sim -1$. The Hubble parameter ($H$) is a negative-valued increasing function of cosmic time as shown in figure 2. The resulting model is contracting and accelerating.

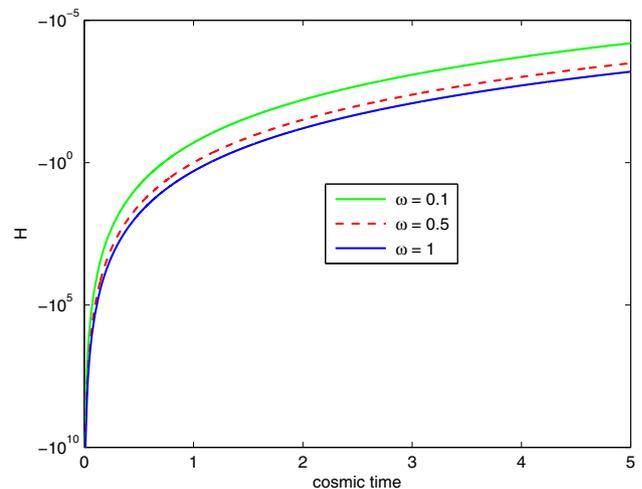

**Figure 2.** Variation of Hubble parameter ($H$) with cosmic time.

For the dust Universe ($\gamma = 1$), $p_m$ for matter is zero during the entire evolution of the Universe. The energy density for the dust Universe is presented in figures 3 and 4. Energy density for matter ($\rho_m$) is a positive-valued decreasing function of time, while $\rho_r$ is negative-valued increasing function of time. At the initial moment ($t = 0$) both are infinite and approach zero for large values of time. The density parameters for matter and radiation are independent of cosmic time and remain constant throughout the evolution of the Universe, as shown by eqs (47) and (48).

For the radiation Universe ($\gamma = \frac{4}{3}$), $p_m$ and $\rho_m$ for matter, $\rho_r$ for radiation, $\Omega_m$ and $\Omega_r$ for matter and radiation are undefined, except total energy density ($\rho$).

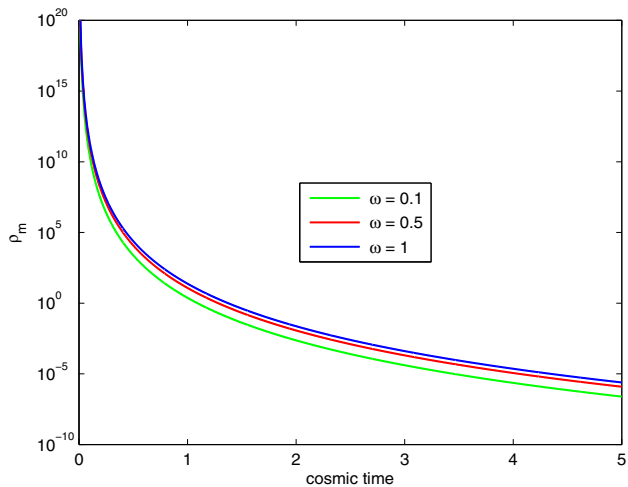

**Figure 3.** Variation of density ($\rho_m$) with cosmic time for $\gamma = 1$.

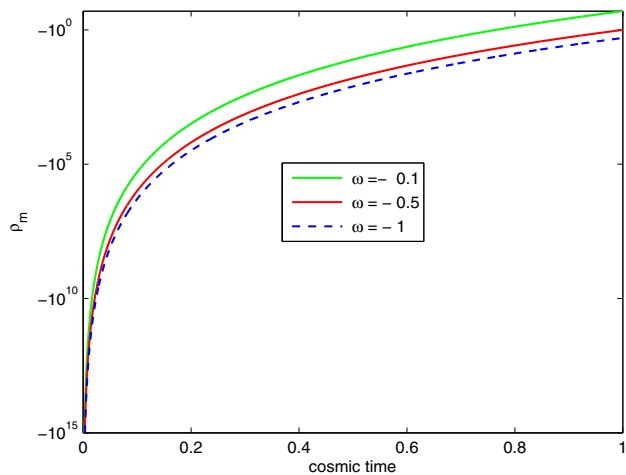

**Figure 6.** Variation of density ($\rho_m$) with cosmic time for $\gamma = 1$.

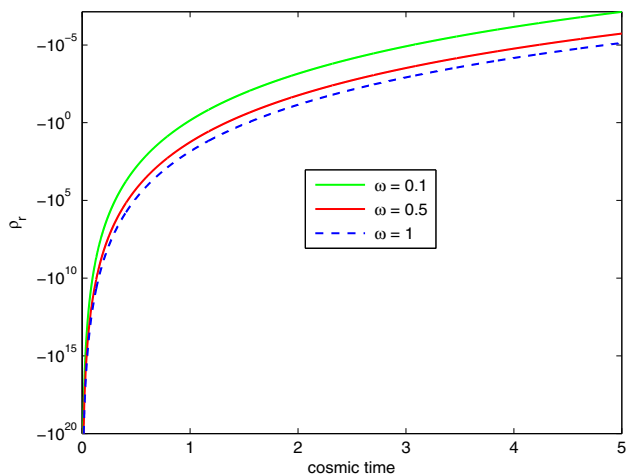

**Figure 4.** Variation of density ($\rho_r$) with cosmic time for $\gamma = 1$.

### 5.2 *The physical behaviour at early time* ($t \to 0$) *and* $\omega < 0$

The scale factor ($S$) evolves from a very small value at the beginning of the Universe to an infinite value at the large cosmic time as shown in figure 5a. Further, it is observed that, when $t = 0$ Hubble parameter is infinite and spatial volume is zero, while spatial volume becomes infinite and Hubble parameter tends to zero for large cosmic time, indicating that the Universe begins with zero volume and explodes at infinite past and future. For large cosmic time, the scalar of expansion becomes zero and initially it is infinite, which shows that the rate of expansion is fast at first and then slows down.

For the dust Universe ($\gamma = 1$), the pressure for matter is zero. Figures 6 and 7 show the energy density for the dust Universe. Energy density for matter is

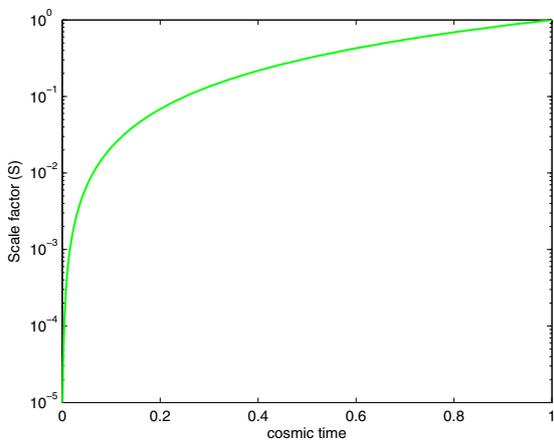

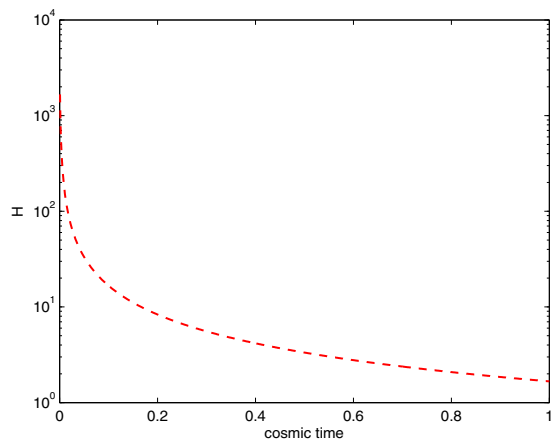

(**a**)                                                (**b**)

**Figure 5.** Variation of (**a**) scale factor ($S$) and (**b**) Hubble parameter ($H$) with cosmic time.

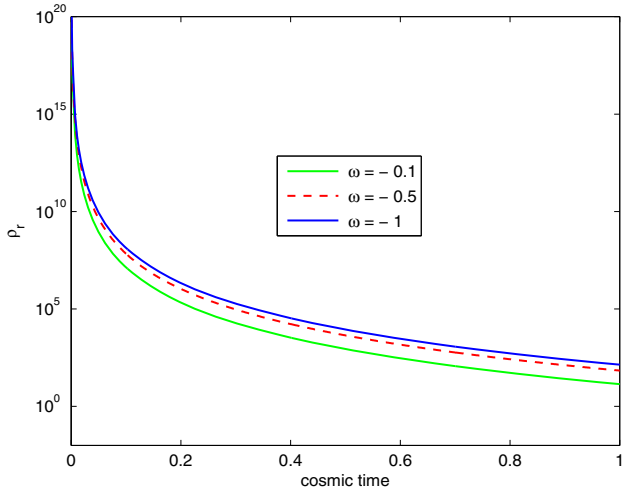

**Figure 7.** Variation of density ($\rho_r$) with cosmic time for $\gamma = 1$.

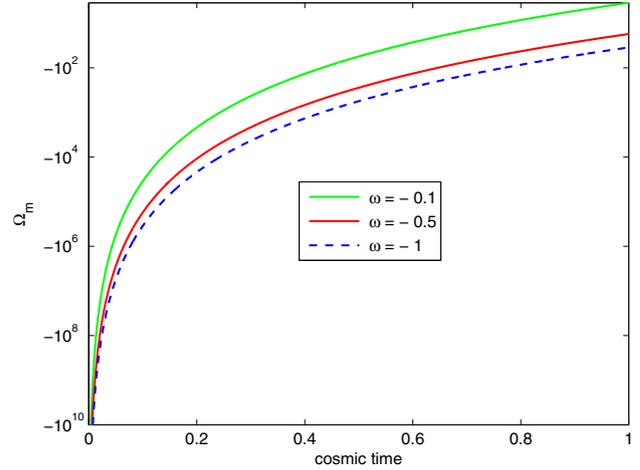

**Figure 8.** Variation of density parameter ($\Omega_m$) with cosmic time for $\gamma = 1$.

negative-valued increasing function of time and energy density for radiation is positive-valued decreasing function of cosmic time. It is observed that initially the Universe is dominated by radiation, which agrees with the actual Universe. From figure 8, we observed that density parameter for matter is negative infinite at $t = 0$, increases with time and finally approaches zero. At the initial epoch ($t = 0$), $\rho_m$ and $\rho_r$ are all infinite. Hence, the model starts evolving with Big-Bang singularity at $t = 0$. As $t \to \infty$, the kinematical and physical parameters of the present model approach zero and spatial volume becomes infinite. Thus, the model essentially gives an empty space for large cosmic time. As the deceleration parameter is negative, the present model represents an accelerated expansion of the Universe which is in good agreement with the recent observations like SNe Ia [33] and CMB anisotropy [34]. The present model leads to the dark energy era. The mysterious dark energy is responsible for the expansion of the Universe. According to the current cosmological observational data, the Universe is dominated by dark energy, which accounts for 70% of the total and 30% dark matter. Here it is important to note that, the results obtained in this model are identical with the results earlier obtained by Samanta [8] in the absence of variable cosmological constant and Adhav *et al* [10].

For the radiation Universe ($\gamma = 4/3$), $p_m$ and $\rho_m$, $\rho_r$, $\Omega_m$ and $\Omega_r$ are undefined, except total energy density $\rho$. The total energy density is infinite at $t = 0$ and tends to zero as $t \to \infty$ (figure 9).

### 5.3 *The physical behaviour at late time* ($t \to \infty$) *and* $\omega > 0$

The scale factor ($S$) decreases as time passes. It is initially infinite at $t = 0$ and approaches zero at a later time,

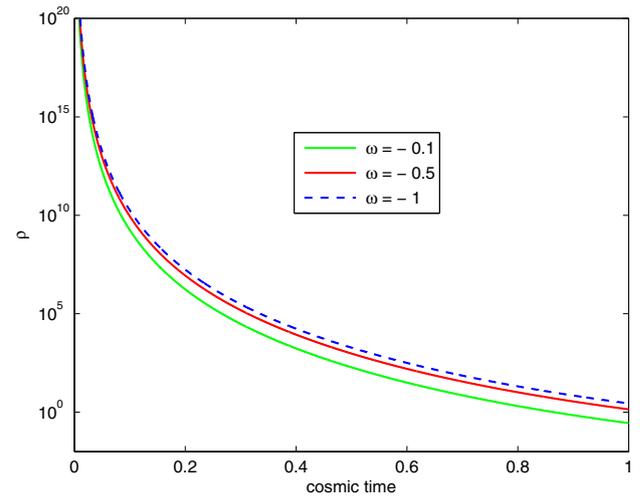

**Figure 9.** Variation of total density ($\rho$) with cosmic time.

as shown in figure 10a. From figure 10b, it is observed that Hubble parameter will be negative infinite at $t = 0$ and tends to zero as $t \to \infty$. It results into the contraction of Universe. From eq. (57), it is observed that initially, spatial volume is infinite and as time increases the Universe approaches a zero volume. Throughout the evolution of the Universe, the deceleration parameter remains constant and negative as observed from eq. (58).

For the dust Universe ($\gamma = 1$), the pressure for matter is zero. From figures 11 and 12 it has been observed that the behaviour of $\rho_m$ is just opposite to that of $\rho_r$. $\rho_m$ is positive and $\rho_r$ is negative and both tend to zero for large cosmic time. Equations (63) and (64) show that, the density parameters $\Omega_m$ and $\Omega_r$ are constant throughout the evolution of the Universe.

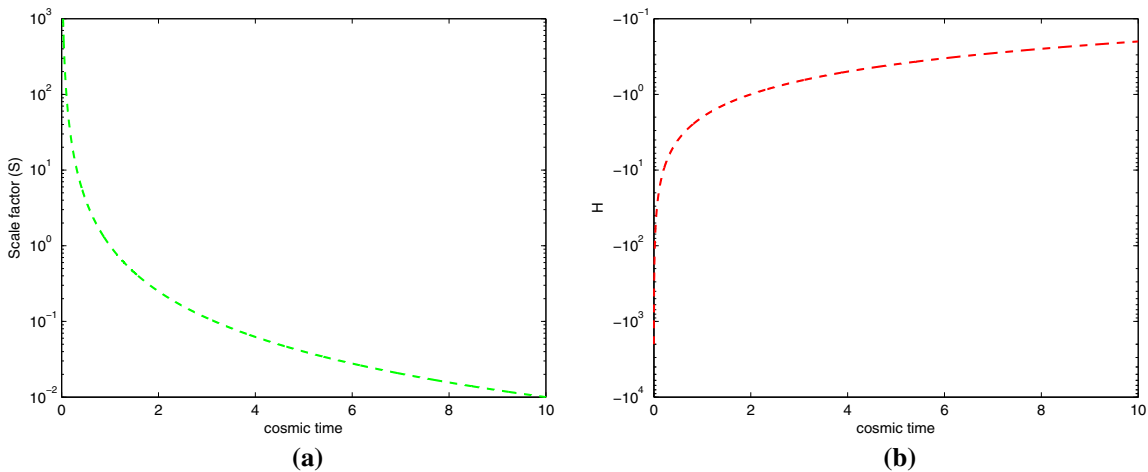

**Figure 10.** Variation of (**a**) scale factor ($S$) and (**b**) Hubble parameter ($H$), with cosmic time.

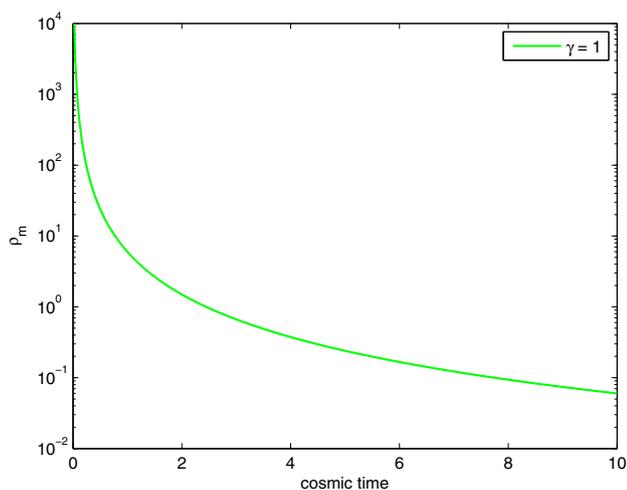

**Figure 11.** Variation of density ($\rho_m$) with cosmic time for $\gamma = 1$.

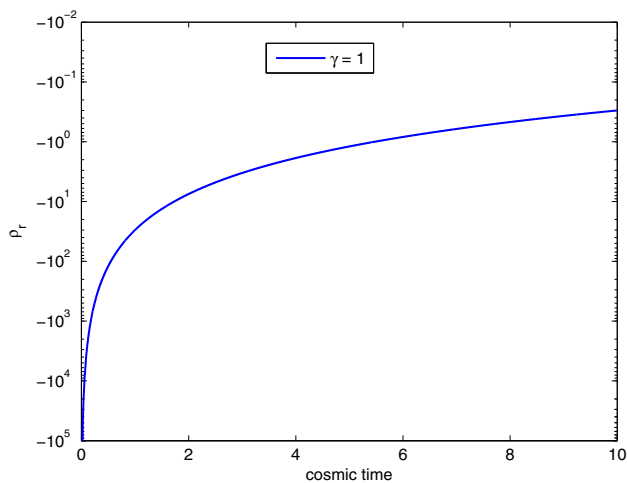

**Figure 12.** Variation of density ($\rho_r$) with cosmic time for $\gamma = 1$.

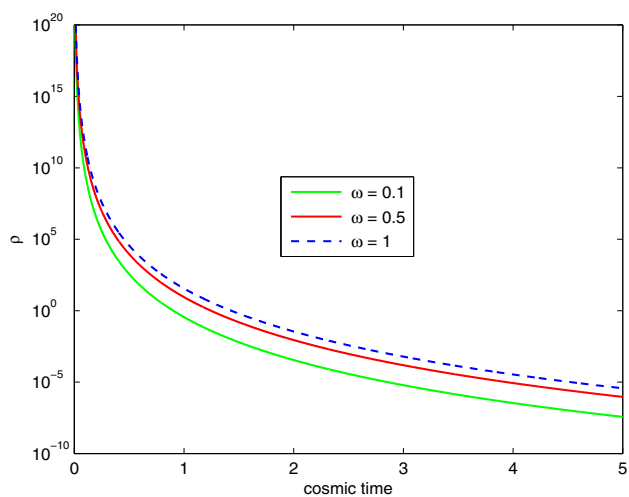

**Figure 13.** Variation of total density ($\rho$) with cosmic time.

For the radiation Universe ($\gamma = \frac{4}{3}$), $p_m$ and $\rho_m$, $\rho_r$, $\Omega_m$ and $\Omega_r$ are undefined, except the total energy density $\rho$.

## 6. Conclusions

In this paper, we have investigated the flat Friedmann–Robertson–Walker two-fluid cosmological models in fractal theory of gravitation. To obtain the solution, we assumed the fractal parameter $\omega \neq 0$ and also considered the relationship between $p_m$ and $\rho_m$, i.e. $p_m = (\gamma - 1)\rho_m$, $1 \leq \gamma \leq 2$. The computations are carried out in the absence of a cosmological constant ($\Lambda = 0$) and within the limits of the UV regime ($\beta = 2$). The scale factor and Hubble parameter diverge at the beginning and vanish for large cosmic time. For the dust Universe, energy densities for matter and radiation approach zero for large values of cosmic time and diverge at the initial moment. The cosmic time has no effect on the density parameters of matter and radiation in Models I and III, while in Model II, density parameters of matter and radiation are functions of cosmic time. In all the models, the total density is a decreasing function of cosmic time and tends to zero for large time (figures 9, 10, 14). In all the investigated models, we found that for the radiation Universe, pressure and energy density for matter and radiation, density parameters are undefined. The Universe is dominated by radiation initially for $\gamma = 1$ in Model II, which agrees with the actual Universe. For the Zeldovich Universe, we observed that densities for matter and radiation have the same behaviour as in Models I and III. The Universe in Model II is expanding and accelerating, with the expansion beginning with the Big-Bang singularity at $t = 0$. Model II is physically meaningful as the associated parameters behave responsibly. The key finding of this paper is that the fractal parameter $\omega$ has a significant effect on the physical and kinematical quantities.

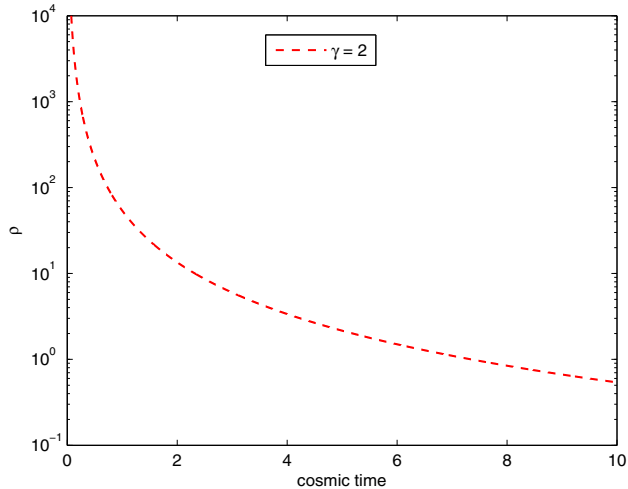

**Figure 14.** Variation of total density ($\rho$) with cosmic time for $\gamma = 1$.

## References

[1] A A Coley and B O J Tupper, *J. Math. Phys.* **27**, 406 (1986)

[2] W Davidson, *Mon. Not. R. Astron. Soc.* **124**, 79 (1962)

[3] C B G McIntosh, *Mon. Not. R. Astron. Soc.* **140**, 461 (1968)

[4] R Fabbri, I Guidi and G Infrarasso, *Proceedings of Second Marcel Grossman Meeting on General Relativity* (Amsterdam, North Hallond, 1992) Vol. 889

[5] G K Goswami and Anirudh Pradhan, 1907.11930v1 [gr-qc]

[6] V U M Rao, G Suryanarayana and Y Aditya, *Adv. Astrophys.* **1**, 1 (2016)

[7] Luis P Chimento, *Phys. Lett. B* **633**, 9 (2006)

[8] G C Samanta, *Int. J. Theor. Phys.* **52**, 11 (2013)

[9] D D Pawar and Y S Solanke, *Int. J. Math. Appl.* **989**, 6 (2018)

[10] K S Adhav, S M Borikar, M S Desale and R B Raut, *Electron. J. Theor. Phys.* **8**, 319 (2011)

[11] D D Pawar and V R Patil, *Prespacetime J.* **4**, 312 (2013)

[12] D D Pawar, S W Bhaware and A G Deshmukh, *Int. J. Theor. Phys.* **47**, 599 (2008)

[13] D D Pawar and A G Deshmukh, *Bulg. J. Phys.* **37**, 56 (2010)

[14] B Mishra, P K Sahoo and Pratik P Ray, 1702.06834v1 [gr-qc] (2017)

[15] S D Katore and S P Hatkar, *Indian J. Phys.* **90**, 243 (2016)

[16] Koijam Manihar Singh, Kangujam Priyokumar Singh and Thiyam Jairam Singh, *Res. Astron. Astrophys.* **11**, 271 (2011)

[17] R K Tiwari, A Beesham and B K Shukla, *Eur. Phys. J. Plus* **132**, 126 (2017)

[18] R K Tiwari, A Beesham and B K Shukla, *Int. J. Geom. Meth. Mod. Phys.* **15**, 1850189 (2018)

[19] V J Dagwal and D D Pawar, *Mod. Phys. Lett. A* **35**, 1950357 (2019)

[20] B Mishra, Fakhereh Md Esmaeili, Pratik P Ray and S K Tripathy, *Phys. Scr.* **132**, 126 (2021)

[21] A H Hasmani and D N Pandya, *Math. Today* **33**, 35 (2017)

[22] B Mandelbrot, *The fractal geometry of nature* (W.H. Freeman and Co., New York, 1982)

[23] L Pietronero, *Physica A* **144**, 257 (1987)

[24] Nottale Laurent, *Int. J. Mod. Phys. A* **7**, 4899 (1992)

[25] A D Linde, *Phys. Lett. B* **175**, 395 (1986)

[26] G Calcagni, *J. High Energy Phys.* **3**, 120 (2010)

[27] Mustafa Salti, Murat Korunur and Irfan Acikgoz, *Eur. Phys. J. Plus* **129**, 1 (2014)

[28] K Karami, Mubasher Jamil, S Ghaffari and K Fahimi, *Can. J. Phys.* **91**, 770 (2013)

[29] H Hossienkhani, H Yousefi and N Azimi, *Int. J. Geom. Meth. Mod. Phys.* **15**, 1850200 (2018).

[30] Ayman A Aly and M M Selim, *Eur. Phys. J. Plus* **130**, 164 (2015)

[31] D D Pawar, D K Raut and W D Patil, *Int. J. Mod. Phys. A* **35**, 2050072 (2020)

[32] S Ghaffari, E Sadri and A H Ziaie, *Mod. Phys. Lett. A* **35**, 2050107 (2020)

[33] A G Riess *et al*, *Astron. J.* **116**, 1009 (1998)

[34] C L Bennet *et al*, *Astrophys. J. Suppl.* **148**, 1 (2003)